\documentclass[conference]{IEEEtran}
\IEEEoverridecommandlockouts

\usepackage{amsmath,amssymb,amsfonts}

\usepackage{algorithmic}
\usepackage{booktabs}
\usepackage{listings}
\usepackage{longtable}
\usepackage{tabularx}
\usepackage{array}
\usepackage{enumitem} 

\usepackage{graphicx}
\usepackage{tikz}
\usetikzlibrary{shapes.geometric, arrows, positioning}

\usepackage{xcolor}
\usepackage{textcomp}
\usepackage{caption}

\usepackage{cite}
\usepackage{url}

\usepackage[margin=1in]{geometry}

\usepackage{hyperref}
\usetikzlibrary{shapes.geometric, arrows, positioning}

\tikzstyle{startstop} = [rectangle, rounded corners, minimum width=3.8cm, minimum height=1cm, text centered, draw=black, fill=blue!20]
\tikzstyle{process} = [rectangle, minimum width=4.2cm, minimum height=1cm, text centered, draw=black, fill=orange!30]
\tikzstyle{io} = [trapezium, trapezium left angle=70, trapezium right angle=110, minimum width=4cm, minimum height=1cm, text centered, draw=black, fill=cyan!20]
\tikzstyle{decision} = [diamond, aspect=2, minimum width=3.5cm, minimum height=1cm, text centered, draw=black, fill=yellow!20]
\tikzstyle{human} = [rectangle, minimum width=4.8cm, minimum height=1cm, text centered, draw=black, fill=purple!20]
\tikzstyle{arrow} = [thick,->,>=stealth]

\definecolor{codegreen}{rgb}{0,0.6,0}
\definecolor{codegray}{rgb}{0.5,0.5,0.5}
\definecolor{codepurple}{rgb}{0.58,0,0.82}
\definecolor{backcolour}{rgb}{0.97,0.97,0.95}

\lstdefinestyle{verilog}{
    backgroundcolor=\color{backcolour},   
    commentstyle=\color{codegreen},
    keywordstyle=\color{blue},
    numberstyle=\tiny\color{codegray},
    stringstyle=\color{codepurple},
    basicstyle=\ttfamily\footnotesize,
    breakatwhitespace=false,         
    breaklines=true,                 
    captionpos=b,                    
    keepspaces=true,                 
    numbers=left,                    
    numbersep=5pt,                  
    showspaces=false,                
    showstringspaces=false,
    showtabs=false,                  
    tabsize=2,
    language=Verilog
}

\def\BibTeX{{\rm B\kern-.05em{\sc i\kern-.025em b}\kern-.08em
    T\kern-.1667em\lower.7ex\hbox{E}\kern-.125emX}}
\begin{document}

\title{Architect in the Loop Agentic Hardware Design and Verification}

\author{\IEEEauthorblockN{Mubarek Mohammed} 
\IEEEauthorblockA{\textit{Western Washington University } \\
\textit{Bellingham, WA USA}\\
mohammm6@wwu.edu}

}

\maketitle

\begin{abstract}
The ever increasing complexity of the hardware design process demands improved hardware design and verification methodologies. With the advent of generative AI various attempts have been made to automate parts of the design and verification process. Large language models (LLMs) as well as specialized models generate hdl and testbenches for small components, having a few leaf level components. However, there are only a few attempts to automate the entire processor design process. Hardware design demands hierarchical and modular design processes. We utilized this best practice systematically and effectively. We propose agentic automated processor design and verification with engineers in the loop. The agent with optional specification tries to break down the design into sub-components, generate HDL and cocotb tests, and verifies the components involving engineer guidance, especially during debugging and synthesis. We designed various digital systems using this approach. However, we selected two simple processors for demonstration purposes in this work. The first one is a LEGv8 like a simple processor verified, synthesized and programmed for the DE-10 Lite FPGA. The second one is a RISC-V like 32-bit processor designed and verified in similar manner and synthesized. However, it is not programmed into the DE-10 Lite. This process is accomplished usually using around a million inference tokens per processor, using a combination of reasoning (e.g gemini-pro) and non-reasoning models (eg. gpt-5-mini) based on the complexity of the task. This indicates that hardware design and verification experimentation can be done cost effectively without using any specialized hardware. The approach is scalable, we even attempted system-on-chip, which we want to experiment in our future work.
\end{abstract}


\section{Introduction}
The domain of digital systems design, traditionally characterized by long development cycles, 
extensive manual effort, and a high barrier to entry, is poised for transformation by advances 
in Artificial Intelligence \cite{fu2024chip}. Large Language Model (LLM) based AI agents now 
possess the capability to understand natural language specifications and generate hardware 
description languages (HDLs) like SystemVerilog and VHDL. This promises to automate tedious 
coding tasks, lower the barrier to entry for hardware design, and accelerate the critical 
design verification loop.

However, moving from single file code generation to agentic, system-level design introduces 
significant challenges related to maintaining architectural integrity, managing complex 
dependencies, and ensuring semantic consistency across dozens of files. Recent work such as 
AutoChip \cite{liu2023autochip} and other end-to-end synthesis frameworks \cite{li2024towards} 
have made significant strides, particularly in high-level synthesis (HLS) and domain-specific 
accelerator design. Yet, a gap remains in understanding the practical capabilities and 
limitations of agentic systems for general-purpose, register-transfer level (RTL) CPU design, 
a domain that demands attention to detail, from instruction set architecture (ISA) 
to microarchitectural data paths.

This case study investigates a practical application of this emerging technology through 
a framework we term the "Blueprint-Driven Agent". We tasked an AI agent with several complete 
design challenges of increasing complexity: a simple single-cycle LEGv8-style processor, 
a synthesizable 32-bit RISC-V core, a pipelined LEGv8 processor, and others(not reported in 
this work). The methodology was rooted in a blueprint-driven approach, where the AI first 
generates a high-level JSON specification, defining the system's architecture, which it then 
uses to create, verify, and self-correct the necessary HDL components by using cocotb testing.

The objective was to rigorously assess the AI's efficacy not just in localized code generation, 
but in its ability to compose a complex, functional, and synthesizable system. While initial 
results in component generation and verification were impressive, system-level verification invariably revealed 
critical failures. This paper documents the collaborative human-AI debugging 
journey, providing a granular analysis of the agent's limitations, and offers a refined 
"Architect-in-the-Loop" workflow as a powerful and economically viable paradigm for future 
hardware design.

\section{Related Works}
In this section we discuss the major works in this area.
\subsection{Verilog Generation and Evaluation}
There are several attempts to fine tune LLMs to better generate Verilog hdl code. Some of them 
are BetterV \cite{betterV}, VerilogEval \cite{VerilogEval}, and RTL coder \cite{RTLCoder}.
The focus of these works is to evaluate the models capability for small modules. 
This is close to the leaf level components or components with few subcomponents. 
Even though we use LLMs to generate SystemVerilog code components, the strength of our work 
comes from how we hierarchically build complex digital systems including processors.

In our work, we mainly use LLMs with a combination of reasoning models (for planning, debugging/fix, 
and integration)  and non-reasoning models (for code and testbench generation).

\subsection{AI Assisted Digital Systems Design}
The work very close to ours conceptually is Chip-Chat \cite{fu2024chip}. They tried to build 
a small processor from scratch by chatting with LLMs. An experienced engineer (the author) 
prompted the LLM to give the required modules, debugged it, and iteratively improved it to 
generate a working processor. Chip-Chat purely involves the engineer and chat bot, with parts 
of the code being broken up into separate chats.They even continued to the tiny-tapeout process for
manufacturing. In our case, we automate the hierarchical and modular design process. 
Another work, AutoChip \cite{liu2023autochip} used LLMs to build small digital circuits by 
incorporating self-feedback loop, very similar to ours at the component level. 
However, AutoChip does not create more complicated components and hierarchical components. 

Our work partially or completely automates the process through agentic digital systems design. 
The experiment in this work is done considering the human engineer playing a critical role 
in the digital systems design process such as a simple processor development process. 
The engineer/architect can provide specification in terms of JSON blueprint as well as a 
text prompt. Once the agent starts, if a component fails the agent initiates a self-correction 
loop. However, it waits for the engineer to provide feedback and approves the fix. The engineer 
at this stage can edit the failed hdl or test files and let the agent continue. In a separate work 
of ours for experimental purposes, not published, we even let the agent run end-to-end without human 
intervention just starting from blueprint or prompt specification. However, even in this case 
the engineer has the option to edit code especially if the agent is struggling to fix some issues.

\section{The AI-Driven Design Methodology}
Our framework is designed to emulate and enhance a modern, specification-driven workflow. It leverages an AI agent for code generation, verification, and self-correction, all under the strategic supervision of a human architect. The core components of this methodology are outlined below.
\begin{figure}[htbp]
\centerline{\includegraphics[width=\columnwidth]{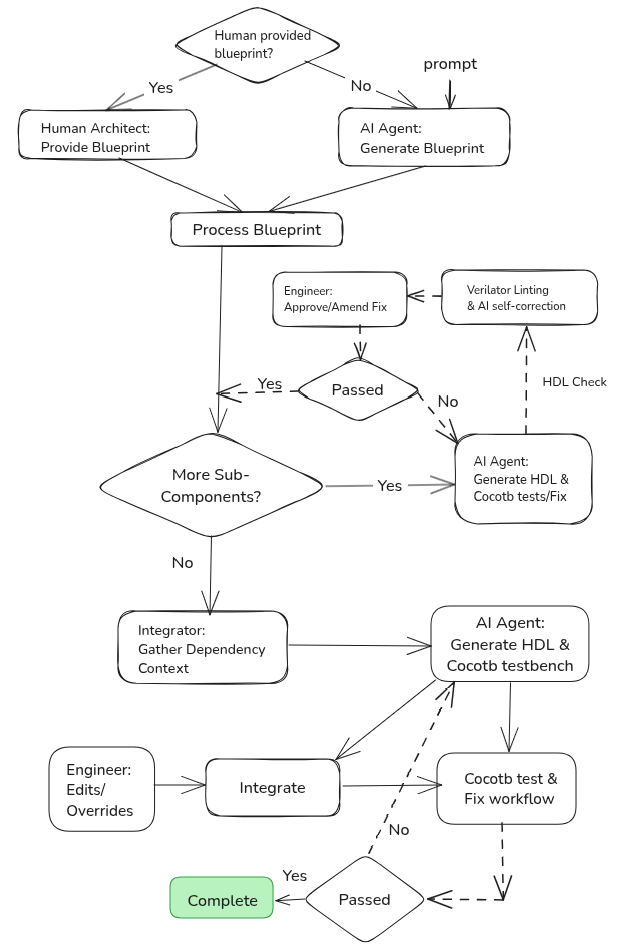}}
\caption{Agentic Digital Design Flow with Human-in-the-
Loop Feedback and Approval at Each Stage}
\label{fig}
\end{figure}

For sample code look at the appendix. The details are also hosted on github.
\footnote[1]{\url{https://github.com/mmubarek/eda/tree/main/paper1}}

\subsection{Phase 1: The Blueprint as a Formal Specification}
\begin{figure}[htbp]
\includegraphics[width=0.8\columnwidth]{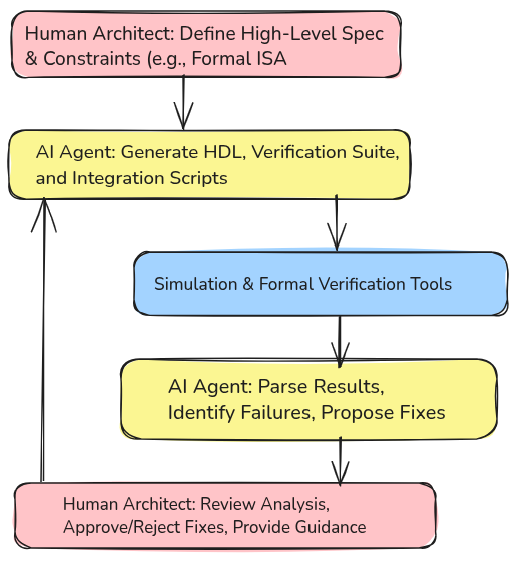}
\caption{Refined Agentic Digital Design Flow with Human-in-the-
Loop}
\label{fig:fig2}
\end{figure}

The foundation of each project is a JSON blueprint, which can either be provided by a human architect or generated by the agent from a high-level natural language prompt. This document serves as the formal, machine-readable specification, defining:
\begin{itemize}[noitemsep,topsep=0pt]
    \item Global project parameters (e.g., data widths, memory depths).
    \item An optional SystemVerilog package for defining a shared Instruction Set Architecture (ISA).
    \item A hierarchical list of all required system components, their individual interfaces (ports and parameters), and their inter-dependencies.
\end{itemize}
A key observation, which became central to later debugging efforts, was that initial agent-generated blueprints often omitted a formal ISA definition or created contradictory parameterizations. Our refined methodology (Figure \ref{fig:fig2}) now employs a highly structured prompt that explicitly guides the agent to produce a complete and non-contradictory blueprint as the first step.

\begin{table}[h!]
    \centering
    \caption{Key Parameters Defined in an Example AI-Generated Blueprint}
    \label{tab:parameters}
    \begin{tabular}{@{}ll@{}}
        \toprule
        Parameter  \\
        \midrule
        \texttt{DATA\_WIDTH}  \\
        \texttt{ADDRESS\_WIDTH}  \\
        \texttt{INSTRUCTION\_WIDTH} \\
        \texttt{OPCODE\_WIDTH}  \\
        \texttt{REG\_ADDR\_WIDTH}  \\
        \bottomrule
    \end{tabular}
\end{table}
In additon to these parameters, ISA related parameters are defined in the blueprint.
\subsection{Phase 2: Component Generation, Integration, and Self-Correction}
Using the blueprint, the agent's orchestrator processes each component according to its dependency graph. For each sub-module, the following workflow is applied:

\begin{enumerate}
    \item \textbf{HDL and Testbench Generation:}  
    The agent generates a synthesizable SystemVerilog module and a corresponding \texttt{cocotb} testbench.  

    \item \textbf{Pre-Flight Linting (Verilator):}  
    A lightweight compilation check using Verilator in linting mode is performed first to efficiently detect syntax and structural errors.  
    \begin{itemize}
        \item If the linting passes, the workflow proceeds to functional verification.  
        \item If the linting fails, the agent analyzes the error log, proposes a fix, and submits it for human approval. Upon approval, the design is retried.  
    \end{itemize}

    \item \textbf{Functional Verification (cocotb):}  
    If linting succeeds, the sub-module is tested against its \texttt{cocotb} unit testbench. Failures here typically expose behavioral or semantic errors.  
    \begin{itemize}
        \item As with linting, testbench failures trigger a self-correction loop: error analysis $\rightarrow$ candidate fix $\rightarrow$ human approval $\rightarrow$ re-verification.  
        \item The number of retry attempts is bounded by a configurable limit (e.g., three attempts per component). If the limit is reached, the human engineer may step in directly to provide manual edits.  
    \end{itemize}

    \item \textbf{Human Overrides:}  
    At any stage, the engineer can manually modify the HDL or testbench. The agent then re-verifies the edited design to ensure correctness before allowing it to proceed.  
\end{enumerate}

This iterative, human-in-the-loop process ensures that every component is both structurally correct and functionally validated before integration.  

Once all sub-modules pass their unit tests, the orchestrator enters the integration phase, generating a top-level module that instantiates and interconnects the verified components. This step is itself subject to the same linting, verification, and human approval loops, since interface mismatches or wiring errors commonly arise during integration.

\subsection{Phase 3: System-Level Verification}
With all components integrated, the agent generates a cycle-accurate top-level \texttt{cocotb} testbench. The verification strategy follows a \emph{golden model} approach: a Python implementation of the ISA maintains an internal model of the processor’s architectural state (registers and memory). For each executed instruction, the golden model predicts the expected state updates, which are then compared against the DUT’s debug outputs at every cycle.  

This strategy uncovers subtle, cross-module issues such as pipeline hazards, bus arbitration errors, or timing mismatches that are not visible at the unit-test level. Failures again trigger the same AI self-correction loop with mandatory human approval, ensuring both automation and accountability.  

Only after passing system-level verification is the design returned to the engineer for FPGA synthesis, hardware debugging, and physical validation. This final step ensures that ultimate responsibility and fine-grained optimization remain in human hands, even as the agent automates the majority of the design flow.

\section{Debugging and Root Cause Analysis}
Debugging and analysis continues for the top level module if there is failure.
This initiates an iterative debugging process that reveals a pattern of agent failure modes. This process was repeated across other designs, confirming our analysis. Table \ref{tab:debuglog} provides a detailed log of representative bugs discovered, their root cause, and the corresponding solution, which often involved refining the agent's underlying prompting strategies.

\renewcommand{\arraystretch}{1.3} 
\begin{table}[htbp]
\centering
\caption{Summary of observed errors, root causes, and applied solutions in LLM-assisted hardware design.}
\label{tab:debuglog}
\begin{tabularx}{\columnwidth}{|>{\raggedright\arraybackslash}X|
                                 >{\raggedright\arraybackslash}X|
                                 >{\raggedright\arraybackslash}X|}
\hline
\textbf{Error} & \textbf{Cause} & \textbf{Solution} \\ \hline

Inconsistent ISA implementations &
LLM generated mismatched instruction encodings across modules &
Refined prompting with explicit ISA tables; enforced consistency via shared JSON blueprint \\ \hline

Parameter propagation errors &
Uncoordinated parameter definitions across modules &
Centralized parameter file and enforced inheritance through JSON schema \\ \hline

Incorrect handling of SystemVerilog packages &
Agent omitted or mismatched imports in integrated designs &
Prompted agent to generate package stubs first, then ensured imports during integration. 
Later, used only centeralized parameters. \\ \hline

Misaligned memory interfaces &
Agent-generated load/store modules had width mismatches &
Human-in-the loop debugging and prompt improvement \\ \hline

Clock related errors &
Agent failed to insert synchronization logic for signals &
Human-in-the-loop correction \\ \hline

\end{tabularx}
\end{table}

\section{Analysis of Agentic Capabilities and Limitations}
This case study, in our view, provides a snapshot of the current capabilities and, more importantly, 
the systemic limitations of LLM agents in digital design.

\subsection{Strengths}
\begin{itemize}
    \item \textbf{Syntactic Correctness and Boilerplate Generation}: The agent excelled at 
    generating syntactically correct SystemVerilog and Python. It handled the boilerplate 
    of module definitions, port lists, and testbench structure well, saving significant manual 
    effort.
    \item \textbf{Blueprint Adherence}: It correctly parsed the JSON blueprint to define module 
    names, parameters, and port widths, demonstrating a strong grasp of structured input and 
    the ability to follow a formal specification. In some cases, the blueprint (JSON) is more 
    than 1300 lines of code.
    \item \textbf{Modular Decomposition and Localized Reasoning}: The initial design breakdown 
    into standard, single-responsibility components was excellent engineering practice. 
    Within the context of a single module, the agent demonstrated a strong ability to implement 
    the required logic based on its description.
\end{itemize}

\subsection{Limitations: The Semantic Cohesion Gap}
The core failures of the agent, across all tested designs, stemmed from a single, 
profound limitation: a lack of \textbf{holistic semantic cohesion}. While capable of reasoning about 
a task in isolation (e.g., "write an ALU"), the agent struggles to maintain a consistent, 
system-wide understanding of how components must interact. This "semantic gap" is a known 
challenge in AI-driven code generation and is particularly acute in hardware design due to 
the tight, bit-level coupling between modules. This manifested in several critical ways, 
summarized in Table \ref{tab:critical_errors}.

\begin{table}[htbp]
\caption{Critical Errors Encountered in Final Phases}
\label{tab:critical_errors}
\centering
\renewcommand{\arraystretch}{1.2}
\begin{tabularx}{\columnwidth}{|X|X|X|}
\hline
\textbf{Error Category} & \textbf{Description of Issue} & \textbf{Resolution / Fix} \\ \hline

Missing Initialization File &
Instruction memory lacked a proper \texttt{.hex} file, leading to failed instruction fetch. &
Created and integrated a correctly formatted \texttt{.hex} file, then re-ran verification. \\ \hline

Cascading Data Width Mismatches &
Fixing the missing \texttt{.hex} file exposed hidden inconsistencies in data widths across modules. &
Performed systematic width alignment across modules, updated bus connections, and extended testbench coverage. \\ \hline

Inconsistent ISA Encoding &
ISA instruction encodings were inconsistently implemented across decoder and execution modules. &
Standardized instruction encodings using a single JSON ISA specification, regenerated RTL and testbenches. \\ \hline

\end{tabularx}
\end{table}

\section{Economic and Future Implications}
\subsection{A Highly Cost-Effective Collaboration}
A remarkable aspect of this project was its economic efficiency. The whole design, verification,
and the debugging process for multiple complex processors was completed by utilizing reasoning models 
for planning and integration, and non-reasoning model for test code generation at component level, 
a simple processor can be completed in less than a million tokens. For context, this stands in stark contrast to the significant engineering man-hours and 
 associated costs typically required for such tasks.  
In this framework, the AI agent plays the role of a junior engineer, automating more than 80\% 
of the repetitive coding and first-pass verification. Meanwhile, the human architect remains 
central—providing high-level guidance, validating decisions, and ensuring that the design 
intent is faithfully captured. The result is not only cost-effective but also aligned with 
democratizing access to sophisticated hardware design workflows.  

\subsection{The Path Forward: The Architect-in-the-Loop}
This work strongly suggests that the most promising near-term trajectory for AI in hardware 
design is not complete autonomy, but a sophisticated \emph{Architect-in-the-Loop} model. 
In this paradigm, the human engineer is not simply a reviewer of the AI’s output but an 
active collaborator from the earliest stages of design.  

Crucially, human oversight begins at the \textbf{specification level}: ambiguous or incomplete 
ISAs, architectural parameters, or high-level constraints must be clarified by the architect 
before the agent proceeds. This ensures that the agent operates with a complete and unambiguous 
blueprint. The human architect also provides strategic debugging insight, approves or rejects 
AI-generated fixes, and may directly modify RTL when necessary, with the agent handling 
re-verification.  

Looking ahead, realizing this vision requires next-generation agents to evolve in three key 
directions:
\begin{enumerate}
    \item \textbf{Demand Complete Specifications}: Agents must be trained to detect missing or contradictory elements in high-level specifications and formally request clarification, rather than making assumptions that can cascade into architectural flaws.  
    \item \textbf{Perform Cross-Module Semantic Analysis}: Agents should employ lightweight static and semantic checks to ensure interface compatibility, parameter consistency, and adherence to design rules across the codebase \emph{before} running expensive simulations.  
    \item \textbf{Integrate with Formal Methods}: Beyond simulation, agents should be capable of generating SystemVerilog Assertions (SVA) or formal properties to express inferred design intent, enabling exhaustive verification with formal tools.  
\end{enumerate}

Taken together, these directions position the human architect as the source of design 
intent, while empowering the AI agent to act as a capable, efficient, and verifiable 
collaborator. This balance preserves accountability, reduces cost barriers, and opens a path 
toward broader participation in advanced hardware design.

\section{Conclusion}
The experiment of designing multiple complex processors using a blueprint-driven AI agent was 
a qualified success. It demonstrated the immense potential of this approach to dramatically 
accelerate hardware development by automating the most labor-intensive aspects of RTL and 
testbench generation. The agent proved capable of creating complex, synthesizable designs 
that successfully targetted FPGA hardware.

However, it also starkly highlighted the "semantic cohesion gap" that prevents current models 
from reliably composing complex systems without human oversight. The agent, while proficient 
at localized tasks, lacks the holistic, system-wide reasoning that is the hallmark of an 
experienced engineer.

The most valuable takeaway is the demonstrated viability and economic efficiency of the 
human-AI collaborative model. The AI acts as a tireless, fast, and exceptionally low-cost 
junior engineer, while the human provides the architectural wisdom and strategic debugging 
insight that is currently irreplaceable. The future of AI in digital design is not about 
replacing the architect, but about providing them with the most powerful co-pilot, 
turning weeks of work into days, and days into hours.

\section*{}

\appendix
\section*{Complete Supporting Data}
You will find the complete blueprint, hdl files, cocotb test, scripts to simulate the digital system at 
\url{https://github.com/mmubarek/eda/tree/main/paper1}. Sample code is provided in the next appendices.

\section*{Appendix: Tools and Debugging Process}

\subsection{Tools}
For simulation, we utilized \texttt{cocotb} together with \texttt{verilator}. We developed both component-level and top-level \texttt{cocotb} tests to ensure functionality and correctness across the design hierarchy.

\subsection{Debugging Process: Synthesis and Experimentation}
One of the major issues encountered was the absence of a hex file for the instruction memory. To resolve this, we created the required hex file, integrated it into the design, and retested the code.  

When compiling for the FPGA, we also needed to provide a dedicated top-level module that wired the design components to the FPGA peripherals. This modification introduced cascading issues, primarily related to mismatched data widths across modules, which required significant debugging. After carefully resolving these mismatches, both verification and synthesis completed successfully.  

The updated design was then iteratively tested using \texttt{cocotb}: each modification was rerun, verified, and debugged through simulation until stable results were achieved.

\subsection{Processor Development Notes}
Using our HDL generation tool, we created a 32-bit RISC-V processor. During early debugging 
stages, we noticed recurring issues. Both the RISC-V and LEGv8 implementations successfully 
passed \texttt{cocotb} tests; however, neither included a preloaded hex file in their 
instruction memory. Instead, \texttt{cocotb} injected instructions directly at the top-level 
testbench.  

To enable standalone execution, we manually added an \texttt{instruction.hex} file for the 
RISC-V design. With this addition, the RISC-V processor ran without errors, though the 
synthesis tool still reported minor warnings.

\section*{Blueprint}
\begin{lstlisting}[language=Verilog, caption=Simple Legv8 Like Blueprint ]
{
  "projectName": "Legv8SingleCycleProcessor",
  "parameters": {
    "DATA_WIDTH": 8,
    "ADDRESS_WIDTH": 8,
    "INSTRUCTION_WIDTH": 16,
    "REG_ADDR_WIDTH": 3,
    "OPCODE_WIDTH": 4,
    "ALU_OP_WIDTH": 3,
    "IMMEDIATE_WIDTH": 6,
    "JUMP_ADDR_WIDTH": 12,
    "PC_INCREMENT_VAL": 2
  },
  "components": [
    {
      "name": "ProgramCounter",
      "file": "program_counter.sv",
      "description": "Holds the address of the current instruction and updates it based on control signals.",
      "dependencies": [],
      "status": "Validating",
      "interface": [
        {
          "name": "clk",
          "direction": "input",
          "width": 1
        },
        {
          "name": "rst",
          "direction": "input",
          "width": 1
        },
        {
          "name": "pc_next_addr",
          "direction": "input",
          "width": "ADDRESS_WIDTH"
        },
        {
          "name": "pc_out",
          "direction": "output",
          "width": "ADDRESS_WIDTH"
        }
      ]
    },
.
.
.
{
      "name": "Legv8SingleCycleProcessor",
      "file": "legv8_single_cycle_processor.sv",
      "description": "Top-level module for the simple single-cycle Legv8 processor, integrating all sub-components.",
      "dependencies": [
        "ProgramCounter",
        "InstructionMemory",
        "ControlUnit",
        "RegisterFile",
        "ALU",
        "DataMemory",
        "SignExtender",
        "Mux2to1",
        "Mux3to1",
        "Adder"
      ],
      "status": "Pending",
      "interface": [
        {
          "name": "clk",
          "direction": "input",
          "width": 1
        },
        {
          "name": "rst",
          "direction": "input",
          "width": 1
        },
        {
          "name": "debug_pc_out",
          "direction": "output",
          "width": "ADDRESS_WIDTH"
        },
        {
          "name": "debug_instruction_out",
          "direction": "output",
          "width": "INSTRUCTION_WIDTH"
        },
        {
          "name": "debug_alu_result",
          "direction": "output",
                "width": "DATA_WIDTH"
        },
        {
          "name": "debug_reg_write_data",
          "direction": "output",
          "width": "DATA_WIDTH"
        }
      ]
    }
  ]
}
\end{lstlisting}

\section{Final Corrected Sample SystemVerilog and Cocotb Test}

\subsection{RegisterFile.sv}
\begin{lstlisting}[language=Verilog, caption=Corrected RegisterFile]
module RegisterFile (
  input logic clk,
  input logic rst,
  input logic [2:0] read_addr1,
  input logic [2:0] read_addr2,
  input logic [2:0] write_addr,
  input logic [7:0] write_data,
  input logic write_en,
  output logic [7:0] read_data1,
  output logic [7:0] read_data2
);

  localparam DATA_WIDTH = 8;
  localparam REG_ADDR_WIDTH = 3;
  localparam NUM_REGISTERS = (1 << REG_ADDR_WIDTH);

  logic [DATA_WIDTH-1:0] registers [NUM_REGISTERS-1:0];

  // Synchronous write, preventing writes to R0
  always_ff @(posedge clk or posedge rst) begin
    if (rst) begin
      for (int i = 0; i < NUM_REGISTERS; i++) begin
        registers[i] <= {DATA_WIDTH{1'b0}};
      end
    end else begin
      // Only write if write_en is high AND the destination is not R0
      if (write_en && (write_addr != 0)) begin
        registers[write_addr] <= write_data;
          end
    end
  end

  // Asynchronous read, forcing R0 to be 0
  // If reading address 0, output 0, otherwise output the register value.
  assign read_data1 = (read_addr1 == 0) ? {DATA_WIDTH{1'b0}} : registers[read_addr1];
  assign read_data2 = (read_addr2 == 0) ? {DATA_WIDTH{1'b0}} : registers[read_addr2];

endmodule
\end{lstlisting}

\subsection{test\_RegisterFile.sv}
\begin{lstlisting}[language=Verilog, caption=Corrected test\_RegisterFile]
import cocotb
from cocotb.clock import Clock
from cocotb.triggers import RisingEdge, FallingEdge, Timer, ClockCycles
import random

@cocotb.test()
async def register_file_test(dut):
    """Test the RegisterFile module thoroughly."""

    # Parameters derived from the HDL module
    DATA_WIDTH = 8
    REG_ADDR_WIDTH = 3
    NUM_REGISTERS = 1 << REG_ADDR_WIDTH # 2^3 = 8 registers

    # Create a clock instance and start it
    clock = Clock(dut.clk, 10, units="ns") # 10ns period = 100MHz clock
    cocotb.start_soon(clock.start())

    # Keep a shadow model of the register file for verification
    # Initialize all shadow registers to 0, matching the reset behavior
    shadow_registers = [0] * NUM_REGISTERS

    cocotb.log.info("Starting RegisterFile test")

    # 1. Reset the DUT
    cocotb.log.info("Applying reset and verifying initial state.")
    dut.rst.value = 1
    dut.write_en.value = 0
    dut.write_addr.value = 0
    dut.write_data.value = 0
    dut.read_addr1.value = 0
    dut.read_addr2.value = 0

    await ClockCycles(dut.clk, 5) # Hold reset for a few clock cycles
    dut.rst.value = 0
    await RisingEdge(dut.clk) # Wait for reset to de-assert and first clock edge

    # Verify all registers are 0 after reset
    for i in range(NUM_REGISTERS):
        dut.read_addr1.value = i
        dut.read_addr2.value = i # Check both read ports
        await Timer(1, units="ns") # Allow combinatorial logic to settle
        assert dut.read_data1.value == 0, \
            f"ERROR: Register {i} (read_data1) not 0 after reset. Got {dut.read_data1.value}"
        assert dut.read_data2.value == 0, \
            f"ERROR: Register {i} (read_data2) not 0 after reset. Got {dut.read_data2.value}"
        shadow_registers[i] = 0 # Ensure shadow model is also reset to 0

    cocotb.log.info("Initial state verified: All registers are 0.")
    
    

    # 2. Basic Write and Read Test
    cocotb.log.info("Performing basic write and read test.")
    test_addr = 1
    test_data = 0xAA
    dut.write_en.value = 1
    dut.write_addr.value = test_addr
    dut.write_data.value = test_data
    await RisingEdge(dut.clk) # Write occurs on this rising edge
    shadow_registers[test_addr] = test_data # Shadow model updated *after* the clock edge

    # Read immediately after write (within the same clock cycle)
    dut.read_addr1.value = test_addr
    dut.read_addr2.value = test_addr
    await Timer(1, units="ns") # Allow combinatorial read outputs to settle
    assert dut.read_data1.value == test_data, \
        f"ERROR: Read data1 mismatch after write. Expected {hex(test_data)}, got {hex(dut.read_data1.value)}"
    assert dut.read_data2.value == test_data, \
        f"ERROR: Read data2 mismatch after write. Expected {hex(test_data)}, got {hex(dut.read_data2.value)}"

    # Read from an unwritten address (should still be 0)
    unwritten_addr = (test_addr + 1) % NUM_REGISTERS
    dut.read_addr1.value = unwritten_addr
    await Timer(1, units="ns")
    assert dut.read_data1.value == shadow_registers[unwritten_addr], \
        f"ERROR: Read data from unwritten register {unwritten_addr} mismatch. Expected {hex(shadow_registers[unwritten_addr])}, got {hex(dut.read_data1.value)}"
    cocotb.log.info(f"Basic write to R{test_addr} ({hex(test_data)}) and read verified.")
     # 3. Concurrent Writes and Reads (different addresses)
    cocotb.log.info("Testing concurrent writes and reads.")
    write_addr1 = 1
    write_data1 = 0x11
    write_addr2 = 2
    write_data2 = 0x22

    # Write to addr1
    dut.write_en.value = 1
    dut.write_addr.value = write_addr1
    dut.write_data.value = write_data1
    await RisingEdge(dut.clk)
    shadow_registers[write_addr1] = write_data1

    # Write to addr2
    dut.write_en.value = 1
    dut.write_addr.value = write_addr2
    dut.write_data.value = write_data2
    await RisingEdge(dut.clk)
    shadow_registers[write_addr2] = write_data2

    # Read both concurrently using different read ports
    dut.read_addr1.value = write_addr1
    dut.read_addr2.value = write_addr2
    await Timer(1, units="ns")
    assert dut.read_data1.value == write_data1, \
        f"ERROR: Concurrent read data1 mismatch. Expected {hex(write_data1)}, got {hex(dut.read_data1.value)}"
    assert dut.read_data2.value == write_data2, \
        f"ERROR: Concurrent read data2 mismatch. Expected {hex(write_data2)}, got {hex(dut.read_data2.value)}"
    cocotb.log.info(f"Concurrent writes to R{write_addr1} ({hex(write_data1)}) and R{write_addr2} ({hex(write_data2)}) and reads verified.")

    # 4. Overwriting a Register
    cocotb.log.info("Testing overwriting a register.")
    overwrite_addr = 2 
    overwrite_data = 0xCC
    dut.write_en.value = 1
    dut.write_addr.value = overwrite_addr
    dut.write_data.value = overwrite_data
    await RisingEdge(dut.clk)
    shadow_registers[overwrite_addr] = overwrite_data

    dut.read_addr1.value = overwrite_addr
    await Timer(1, units="ns")
    assert dut.read_data1.value == overwrite_data, \
        f"ERROR: Overwrite failed for R{overwrite_addr}. Expected {hex(overwrite_data)}, got {hex(dut.read_data1.value)}"
    cocotb.log.info(f"Overwrite of R{overwrite_addr} with {hex(overwrite_data)} verified.")

    # 5. No Write when write_en is low
    cocotb.log.info("Testing no write when write_en is low.")
    no_write_addr = 3
    no_write_data = 0xFF
    original_value = shadow_registers[no_write_addr] # Should be 0 from reset
    dut.write_en.value = 0 # Crucial: write_en is low
    dut.write_addr.value = no_write_addr
    dut.write_data.value = no_write_data
    await RisingEdge(dut.clk) # Clock edge, but no write should occur

    dut.read_addr1.value = no_write_addr
    await Timer(1, units="ns")
    assert dut.read_data1.value == original_value, \
        f"ERROR: Write occurred when write_en was low for R{no_write_addr}. Expected {hex(original_value)}, got {hex(dut.read_data1.value)}"
    assert dut.read_data1.value != no_write_data, \
        "ERROR: Data was written despite write_en being low!"
    cocotb.log.info(f"No write to R{no_write_addr} when write_en is low verified.")

    # 6. Fill all registers and verify
    cocotb.log.info("Filling all registers and verifying their contents.")
    dut.write_en.value = 1
    for i in range(NUM_REGISTERS):
        data_to_write = (i * 17 + 5) & ((1 << DATA_WIDTH) - 1)
        dut.write_addr.value = i
        dut.write_data.value = data_to_write
        dut.write_en.value = 1 if i != 0 else 0  # Don't write to R0
    
        await RisingEdge(dut.clk)
    
        # Update shadow model only if not R0
        if i != 0:
            shadow_registers[i] = data_to_write
        else:
            shadow_registers[i] = 0  # R0 always stays 0

    # Verify all registers by reading them back
    for i in range(NUM_REGISTERS):
        dut.read_addr1.value = i
        dut.read_addr2.value = (i + 1) % NUM_REGISTERS # Read another one concurrently
        await Timer(1, units="ns")
        assert dut.read_data1.value == shadow_registers[i], \
            f"ERROR: Verification failed for R{i} (read_data1). Expected {hex(shadow_registers[i])}, got {hex(dut.read_data1.value)}"
        assert dut.read_data2.value == shadow_registers[(i + 1) % NUM_REGISTERS], \
            f"ERROR: Verification failed for R{(i + 1) % NUM_REGISTERS} (read_data2). Expected {hex(shadow_registers[(i + 1) % NUM_REGISTERS])}, got {hex(dut.read_data2.value)}"
        cocotb.log.debug(f"Verified R{i} ({hex(shadow_registers[i])}) and R{(i + 1) % NUM_REGISTERS} ({hex(shadow_registers[(i + 1) % NUM_REGISTERS])})")
    cocotb.log.info("All registers filled and verified successfully.")

    # 7. Randomized Test
    cocotb.log.info("Starting randomized test (100 cycles).")
    
    for cycle in range(100):
        # 1. Determine write operation for the *upcoming* clock edge
        write_this_cycle = False
        write_addr_val = 1
        write_data_val = 0

        if random.random() < 0.6: # 60% chance to write
            write_this_cycle = True
            write_addr_val = random.randint(1, NUM_REGISTERS - 1)
            write_data_val = random.randint(1, (1 << DATA_WIDTH) - 1)
            dut.write_en.value = 1
            dut.write_addr.value = write_addr_val
            dut.write_data.value = write_data_val
            cocotb.log.debug(f"Cycle {cycle}: Setting inputs for write: addr={write_addr_val}, data={hex(write_data_val)}")
        else:
            dut.write_en.value = 0 # No write this cycle
            cocotb.log.debug(f"Cycle {cycle}: No write this cycle.")

        # 2. Set read addresses for the *current* cycle's verification.
        # These reads will reflect the state *after* the upcoming clock edge.
        read_addr1 = random.randint(0, NUM_REGISTERS - 1)
        read_addr2 = random.randint(0, NUM_REGISTERS - 1)
        dut.read_addr1.value = read_addr1
        dut.read_addr2.value = read_addr2

        # 3. Wait for the clock edge. This is when the DUT's write (if enabled) happens.
        await RisingEdge(dut.clk)

        # 4. Update shadow model *after* the clock edge if a write was enabled
        if write_this_cycle:
            shadow_registers[write_addr_val] = write_data_val

        # 5. Allow combinatorial logic for reads to settle *after* the clock edge
        await Timer(1, units="ns")

        # 6. Now verify the reads
        expected_data1 = shadow_registers[read_addr1]
        expected_data2 = shadow_registers[read_addr2]
        actual_data1 = dut.read_data1.value
        actual_data2 = dut.read_data2.value

        assert actual_data1 == expected_data1, \
            f"ERROR: Cycle {cycle}: Random Read1 mismatch at addr {read_addr1}. Expected {hex(expected_data1)}, got {hex(actual_data1)}"
        assert actual_data2 == expected_data2, \
            f"ERROR: Cycle {cycle}: Random Read2 mismatch at addr {read_addr2}. Expected {hex(expected_data2)}, got {hex(actual_data2)}"
        cocotb.log.debug(f"Cycle {cycle}: Random Read: addr1={read_addr1}, data1={hex(actual_data1)} (expected {hex(expected_data1)}) | "
                          f"addr2={read_addr2}, data2={hex(actual_data2)} (expected {hex(expected_data2)})")

    cocotb.log.info("Randomized test completed.")
    cocotb.log.info("RegisterFile test finished successfully!")
\end{lstlisting}


\begin{thebibliography}{00}
\bibitem{fu2024chip}
H. Fu, H. G. Gerez, et al. (2024). \textit{Chip-Chat: A Large Language Model for Conversational Hardware Design}. arXiv:2402.13324.
\bibitem{liu2023autochip}
Z. Liu, H. G. Gerez, et al. (2023). \textit{AutoChip: A LLaMA-based Agent for Chip Design}. arXiv:2311.09228.
\bibitem{li2024towards}
B. Li, Y. G. Gerez, et al. (2024). \textit{Towards Automatic and Agile AI/ML Accelerator Design with End-to-End Synthesis}. In Proceedings of the 2024 ACM/SIGDA International Symposium on Field-Programmable Gate Arrays.
\bibitem{betterV}
Z. Pei, et al., "BetterV: Controlled Verilog Generation with Discriminative Guidance," arXiv preprint arXiv:2402.03375, 2024.
\bibitem{RTLCoder} 
S. Liu, et al., "RTLCoder: Outperforming GPT-3.5 in Design RTL Generation with Our Open-Source Dataset and Lightweight Solution," arXiv preprint arXiv:2312.08617, 2023.
\bibitem{VerilogEval} 
M. Liu, N. Pinckney, B. Khailany, and H. Ren, "VerilogEval: Evaluating Large Language Models for Verilog Code Generation," arXiv preprint arXiv:2309.07544, 2023.
\bibitem{cocotb} 
Cocotb Contributors, cocotb documentation, [Online]. Available: https://docs.cocotb.org/. (Accessed: Aug. 26, 2025).
\bibitem{Verilator} 
W. Snyder, Verilator,[Online]. Available: https://github.com/verilator/verilator, 2023 (Accessed: August 26, 2025)

\end{thebibliography}
\end{document}